# Deconstructing the Tail at Scale Effect Across Network Protocols

Akshitha Sriraman   Sihang Liu   Sinan Gunbay   Shan Su   Thomas F. Wenisch

University of Michigan, Ann Arbor

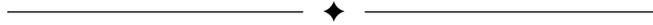

**Abstract**—Network latencies have become increasingly important for the performance of web servers and cloud computing platforms. Identifying network-related tail latencies and reasoning about their potential causes is especially important to gauge application run-time in online data-intensive applications, where the $99^{th}$ percentile latency of individual operations can significantly affect the the overall latency of requests.

This paper deconstructs the "tail at scale" effect across TCP-IP, UDP-IP, and RDMA network protocols. Prior scholarly works have analyzed tail latencies caused by extrinsic network parameters like network congestion and flow fairness. Contrary to existing literature, we identify surprising rare tails in TCP-IP round-trip measurements that are as enormous as 110x higher than the median latency. Our experimental design eliminates network congestion as a tail-inducing factor. Moreover, we observe similar extreme tails in UDP-IP packet exchanges, ruling out additional TCP-IP protocol operations as the root cause of tail latency. However, we are unable to reproduce similar tail latencies in RDMA packet exchanges, which leads us to conclude that the TCP/UDP protocol stack within the operating system kernel is likely the primary source of extreme latency tails.

## 1 INTRODUCTION

Online Data Intensive (OLDI) applications like web search, online retail, and advertising comprise a significant fraction of data center applications [12]. Meeting soft real-time deadlines determines end-user experience and is hence of paramount importance for this application class. For instance, web search operators seek a response time of less than $300ms$ so that search results feel instantaneous to end users [10].

OLDI applications access multi-terabyte data sets while responding to user-facing latency-sensitive queries by dividing/sharding their data sets across hundreds or thousands of servers/leaf nodes. Responses from individual leaf nodes are aggregated to form the final query response [20, 5, 6]. Hence, the response times of individual leaf nodes critically contribute to the overall query response time. Increases in computational performance of modern day processors lead to auxiliary functionalities, such as I/O interactions, network operations, and disk accesses, becoming significant performance bottlenecks. As query response times must be less than $300ms$, even millisecond-scale stalls in constituent operations result in substantial fluctuations in the overall query response time. When scaled up to warehouse-scale computing platforms employed by OLDI applications, such fluctuations inevitably cause some leaf nodes to lag behind others. Since the latency of such massively parallel programs depends on the slowest process, enormous run-time variations often lead to significant performance degradation. This effect is known as the tail latency problem (a.k.a. the tail-at-scale effect) and is a dominant bottleneck in interactive web applications [14, 4, 1, 8, 11].

The tail latency problem is often addressed by replicating requests to leaf servers such that query responses depend on data from only one duplicate request, thereby reducing the probability of waiting for stragglers. Alternatively, an overall query result may be returned even though only a subset of leaf servers have responded; straggling responses are simply ignored. Whereas the former solution is often infeasible due to its prohibitive expense, the latter sacrifices result quality [14] compromising user experience [21, 17, 15]. Moreover, processing only a subset of results may not be universally applicable. For example, when a user is searching their own email for a particular message, providing only a subset of results may be unacceptable. In such cases, identifying and mitigating the true causes of tail latency becomes critical for application performance.

Although high tail latencies experienced by individual elements of a service can be influenced by aspects like global resource sharing, maintenance ac-

. This work was published in the 2016 Workshop on Duplicating, Deconstructing and Debunking (WDDD) held in association with the International Symposium on Computer Architecture (ISCA).

tivities, and energy management [4], the end-to-end network-related latency incurred by a web query is often cited as a major contributor to the tail latency problem [2]. However, prior works attribute performance degrading latency variations to network aspects like network switch congestion [14, 8] and adaptive network packet flow prioritization [9, 10, 11, 13]. In contrast, we show that surprising outlying tail latency in network packet delivery can still arise when we eliminate such commonly blamed causes, such as network congestion or oblivious network routing. High latency fluctuations can be caused by a more fundamental aspect of the network design: the network protocol stack in the operating system kernel. Popular network protocols like Transmission Control Protocol (TCP) and User Datagram Protocol (UDP) over the Internet Protocol (IP) [14] were designed and perfected for an era where CPU processing latencies were formidable enough to mask the sub-millisecond overheads introduced by these protocols. Therefore, contemporary processing latencies suffer from obsolete network protocol design.

Furthermore, it is a common misconception that the TCP-IP protocol necessarily incurs larger tail latencies compared to UDP-IP, given that TCP-IP performs sophisticated operations like TCP windowing, error checking, packet ordering, and three-way handshakes that introduce a higher performance overhead than the simpler UDP-IP [18]. Our work shows that, contrary to popular belief, there is no notable difference between tail latencies incurred in TCP-IP and UDP-IP round-trips under uncongested network conditions, thereby implying that the additional protocol steps in TCP-IP do not explain extreme latency tails.

In this paper, we measure the extent of the tail-at-scale effect in the network layer as observed in the two most common network protocols: TCP and UDP over IP [14]. For each protocol, we document the contribution (or lack thereof) of varying offered load on tail latency. Finally, we compare these protocols against a Remote Direct Memory Access (RDMA) protocol. We observe far more modest tails in RDMA round-trips, leading us to conclude that hiccups or outlying behavior in the TCP-IP and UDP-IP software protocol stack are the most likely source of extreme tails. The key contributions of this paper are:

- We identify surprising sources of tail latency in the network layer that arise from software sources like the network protocol stack in the operating system kernel.
- We discern no conspicuous difference between tail latencies exhibited by TCP-IP and UDP-IP round trips, thereby eliminating the additional TCP-IP protocol operations as the source of extreme latency tails.

The remainder of the paper is organized as follows: We characterize "conventional wisdom" on network latency tails in Section 2 and provide brief background for commonly used networking protocols in Section 3. Section 4 describes our experimental methodology, while Section 5 presents our tail latency evaluations for TCP-IP, UDP-IP, and RDMA protocols. We provide additional discussion in Section 6 and we conclude in Section 7.

## 2 CONVENTIONAL WISDOM

Large variations in query response times are generally caused by factors like resource sharing, background daemons, network queuing, and power management features [4]. End-to-end response time fluctuations caused by the network layer substantially impact the quality of service (QoS) offered by data center applications [9, 14, 10, 13, 11, 8]. Solutions that focus on long tail latencies that are the result of non-network related factors [7, 1], are beyond the scope of this paper.

Prior works that identify the network as a critical tail-inducing component fall into two broad categories, (1) those that prioritize network flows based on the latency sensitivity of applications, and (2) those that seek to manage network congestion.

**Prioritizing network flows based on latency sensitivity of applications.**
Web-based data center applications primarily employ TCP-IP for data transfer [14]. Network protocols that adopt TCP-IP (or its variants [14, 10]) use approximate fair sharing to partition link bandwidth obliviously (and typically equally) among network flows. The network-traffic incognizant nature of such protocols results in considerable degradation of application response time. To this effect, Hong et al. [9] propose a preemptive flow scheduling protocol that completes network flows such that they meet soft real-time deadlines. Moreover, Zats et al. [11] introduce a cross-layer network stack design that uses application-specified flow priorities to provide adaptive load balancing, thereby reducing long tails. Likewise, Zhu et al. [3] recommend a scheme to proactively configure rate limits and application priorities across multiple shared network stages to curtail long tails. Literature that recommends developing deadline-conscious protocols include Wilson et al. [13], who used explicit rate control information to proportionately allocate bandwidth depending on flow deadlines. Similarly, Vamanan et al. [10] designed a deadline-aware TCP-IP protocol that incorporates congestion avoidance and appropriate bandwidth allocation to enable OLDI applications to meet soft real-time deadlines.

**Reducing network congestion.**
On the other hand, Alizadeh et al. [14] attribute network-induced tail latencies to switch queuing delays and propose reducing switch buffer occupancy time by using Explicit Congestion Notification (ECN) to monitor the extent of network congestion. Follow-up literature [8] advocates capping link utilization at less than link capacity to create an allowance for latency-sensitive traffic to avoid being buffered, thereby eliminating large buffer queuing delays. While factors like network congestion and flow fairness are undeniable tail-introducing culprits, we find that, even if these sources are eliminated, astonishing tail latencies still arise. Our experiments indicate surprising sources of large latency variations in the $99.9^{th}$ percentage response time, which we attribute to the software network protocol stack in the operating system kernel.

## 3 BACKGROUND

We consider the network's role in contributing to the tail-at-scale problem. We are particularly interested in deconstructing the tail-at-scale effect across TCP-IP, UDP-IP, and RDMA protocols and identifying probable factors that contribute to the tail. We briefly describe the TCP, UDP, and RDMA protocols.

### 3.1 TCP

Transmission Control Protocol (TCP) is connection-oriented, that is, the source and destination must establish a dedicated connection before data can be transmitted between them. TCP divides a byte-stream into segments and exchanges data at segment granularity. TCP supports re-transmission of lost packets by employing a three-way handshake. It also ensures ordered segment delivery and checks packet integrity.

The data processing path followed by the TCP protocol is indicated by the blue blocks of Fig. 1. Message transmission broadly encompasses 8 steps [19]: (1) the message is copied into the socket buffer; (2) it is then sub-divided into segments; (3) the resulting segments are checked for errors; (4) the segments are then passed to the underlying IP layer; (5) the IP layer extends the segments to create IP datagrams that are then transferred to the device driver; (6) the device driver adds packet descriptors and delivers the packet to the network adapter (NIC); (7) the network adapter performs a DMA operation to move the data indicated by the descriptor from the socket buffer to the NIC buffer; (8) the NIC finally places the data on the physical link and signals an interrupt to announce the completion of segment transmission.

Message reception involves reversing the data path described above by concatenating received segments to obtain the final data. It is vital to note that TCP-IP connections are managed by the operating system through the socket programming interface. The importance of this aspect will become more clear in successive sections.

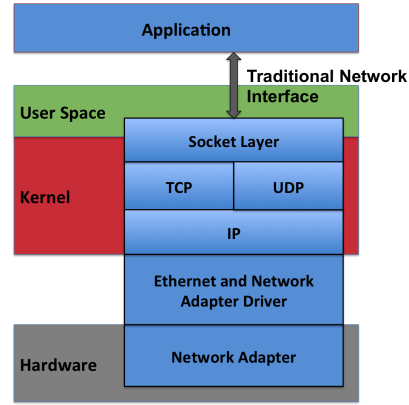

**Fig. 1:** The TCP-IP/UDP-IP Stack

### 3.2 UDP

In contrast to TCP, the User Datagram Protocol (UDP) is a "connectionless" protocol. To ensure fast data transmission, UDP omits handshaking and therefore does not guarantee reliable delivery. Additionally, UDP does not provision for error correction mechanisms or message ordering.

Apart from the protocol nuances mentioned above, the UDP data transmission and reception paths are similar to those of TCP, as shown in Fig. 1. Messages are copied to socket buffer, broken into segments and passed to the IP layer which then transfers messages in the form of datagrams through the physical link.

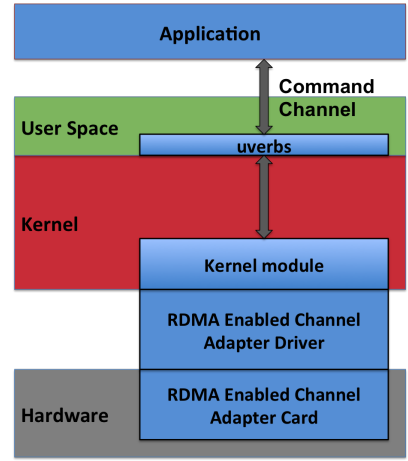

**Fig. 2:** RDMA data channel setup using the command channel

### 3.3 RDMA

Remote Direct Memory Access (RDMA) is a data communication protocol that transfers messages by moving buffers between two applications across the network. The principal difference between RDMA and traditional transport protocols like TCP and UDP lies in the fact that RDMA bypasses the operating system while transferring data. The network adapter directly sends/receives data to/from application memory, thereby eliminating the need to copy data between application memory and data buffers in the operating system. Such transfers offer comparably lower latencies and higher throughput, as they do not require any work from the CPU or caches. Furthermore, the latency incurred due to all operating system activities is eliminated. Therefore, we can use the RDMA protocol as a control in experiments to determine the impact that the operating system protocol stack introduces in tail latencies.

RDMA requires a network adapter card called the Host Channel Adapter (HCA) that enables an RDMA engine. To initiate data transfer, the HCA builds a data channel that extends from the RDMA engine to the application memory over the PCI Express bus. To establish the data channel, RDMA uses a kernel driver to first create a command channel, as shown in Fig. 2. The command channel uses the "verbs" API to establish data channels that are capable of bypassing the operating system kernel while transferring data, as depicted by the blue blocks in Fig. 3.

RDMA augments successful data transfer by using the established data channels to read/write to/from send/receive queues directly. The HCA comprises logic for protocol operations like segmentation and reassembly, flow control, and reliable delivery. Hence, the HCA hardware is responsible for executing the RDMA protocol to enable communication.

The steps involved in transferring data from a client to the server using RDMA are summarized below:

- Initially, memory regions that include the send queue, receive queue, and the completion queue are *registered*, providing the HCA access to the corresponding virtual memory locations.
- The client's HCA streams data from the send queue of the client to the receive queue of the server.
- The server's HCA consumes the data from the receive queue.
- The data thus streams over a high speed data channel, bypassing the operating system kernel.
- When data streaming from the send queue completes, both the client and the server place a completion message in their respective completion queues, thereby indicating success of the data transfer.

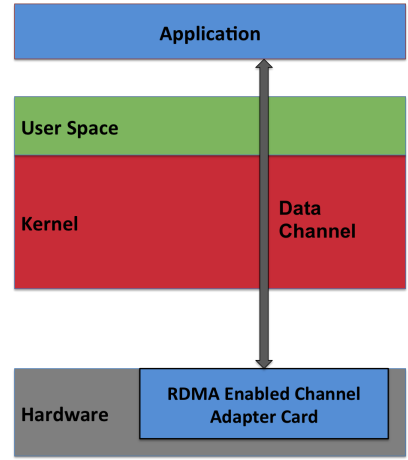

**Fig. 3:** RDMA communication through the established data channel

## 4 Design and Methodology

TCP over Ethernet is the dominant networking channel [14], so understanding the latency distribution and throughput bottlenecks exhibited by TCP-IP can pave the way for exposing network protocol related factors that precipitate tail vulnerabilities in data center applications.

To evaluate the prevalence of the tail latency problem in network protocols, it is imperative to eliminate the effect of less-intrinsic network effects like network switch congestion, packet scheduling, load balancing techniques, and flow prioritization on application performance. On these grounds, we establish an isolated network consisting of two Linux machines with network adaptors directly connected via a cross-over Ethernet cable. This private network organization ensures pristine round trip time (RTT) measurements of network packet communication in a simple server-client interaction. Each network communication consists of the client machine initiating a network connection by sending a data packet to the server and measuring the time it takes for the packet to return to the client. The server is a simple echo program that immediately echos the packet it received back to the original sender.

Typically, recording time sensitive measurements that involve multiple systems entails an explicit need to synchronize clocks across machines to avoid measurement discrepancies. However, we measure time only at the initiator of each round-trip exchange and measure message RTT. One disadvantage of this method is that each RTT data point is then the sum of two network traversals, either or both of which may exhibit a tail behavior. However, we observe that the magnitude of the tails we seek are drastically larger than median RTT latency and the additional network hop can be ignored.

It is no secret that TCP-IP manifests a higher communication overhead than UDP-IP due to additional protocol intricacies [18]. Furthermore, fair sharing of link bandwidth irrespective of worker thread priorities considerably adds to TCP's performance overhead [14]. These performance overheads were deemed significant enough for Facebook to implement customized protocols on top of UDP-IP for several critical services [22]. To determine if extreme tails are a particular symptom of the additional features of TCP-IP, we also perform RTT measurements for UDP-IP. Furthermore, to isolate the effect of the networking hardware from software protocol stacks at each endpoint, we contrast both TCP-IP and UDP-IP against RDMA, wherein the operating system at both the client and remote play no role.

The following sections detail the specific experiments we devised for each protocol:

### 4.1 TCP-IP

We implement a simple client that uses TCP-IP to connect to the server at a single port, prepares data messages of predetermined sizes, and proceeds to log the latency of each message exchange. An exchange begins with the *send()* system call that releases the packet into the network, and ends when a *recv()* system call with a response from the remote returns. For each experiment, we perform $100,000$ exchanges and log their RTTs.

Our TCP-IP experiment models a basic server-client interaction. We further study the impact of varying the message payload size and scenarios where we seek to saturate network bandwidth by transmitting simultaneously from multiple server-client pairs, each listening on a different port. Multiplexing multiple connections in this fashion serves as a stress-test. Our interest is to investigate the degree to which additional traffic makes the tail latency distribution worse. Furthermore, we study the RTT distribution for a non-blocking TCP-IP server-client pair. The goal is to characterize the tail effect when TCP-IP emulates the RDMA protocol's non-blocking nature.

### 4.2 UDP-IP

The client implementation for the UDP-IP protocol does not require explicit connection setup and tear down. Messages of specific sizes are prepared and sent to the target address-port combination using the UDP-IP *send()* and *recv()* system calls. The RTT data is ultimately gathered for $100,000$ such exchanges.

Our baseline UDP-IP experiment again models a simple server-client pair. Similar to our TCP-IP stress-tests, we also investigate multiple concurrent UDP-IP server-client pairs to observe the impact of additional traffic on tail latencies. We again vary message payload size within the range allowed by UDP.

### 4.3 RDMA

Our RDMA experiments are structured differently than our TCP-IP and UDP-IP experiments due to the drastic differences of the verbs API from conventional sockets. Although we seek to model a similar server-client exchange, the RDMA programming model requires reading/writing directly to/from remote memory, asynchronous requests and completion notifications, and non-blocking system calls for data transfer, unlike the other network protocols.

Our server process is designed to create and bind event channels to an address for connection requests, create a listener that waits for connection requests, register a memory region, construct the completion queue and the send-receive queue pair, accept and ensure connection establishment, and post read/write operations through the established connection. The client creates an event channel for address resolution, constructs and binds a connection identifier to the local RDMA device, registers a memory region, constructs send/receive queues and completion queues, performs server route resolution, initiates and waits for connection establishment, and initiates read/write operations over the established queue-pair.

Our RDMA experiments involve posting read/write requests to the client send queue and spawning a thread that polls the completion queue. We time message exchanges from the moment a read/write request is posted to the send queue until the poll loop spinning on the completion queue indicates that a completion notification has arrived. Note that, in the underlying protocol, these operations still represent a network round-trip. We again measure $100,000$ read/write RDMA exchanges.

Our RDMA experiments use a single server-client pair to establish a baseline latency distribution. We then vary the size of read/write requests to understand the impact of payload size on tail latency. As RDMA style communication completely bypasses the operating system, the RDMA experiments allow us to distinguish latency tails in TCP and UDP that are likely attributable to software.

## 5 EVALUATION

We perform experiments on an isolated network comprising two Linux machines. We use Mellanox ConnectX-3 network adapters directly attached via 10 Gbps Ethernet cabling. The specifications of the two Linux systems are listed in Table 1. We use the Linux `gettimeofday` API from the standard "time.h" library as the time source for latency measurements. While collecting RTT samples, no other programs were explicitly launched. We present RTT measurements as the average of 10 runs, after excluding the slowest and fastest runs. Metrics used in the evaluation are defined below.

|  | Client | Server |
|---|---|---|
| CPU | i7 5820K | i7 6700 |
| $\mu$Architecture | Haswell | Skylake |
| #Physical Cores | 6 | 4 |
| Hyperthreading | Y | |
| #Logical Cores | 12 | 8 |
| Clock Freq | 1.2GHz | |
| L1 Cache | 192KB | 128KB |
| L2 Cache | 1.5MB | 1MB |
| L3 Cache | 12MB | 8MB |
| RAM Size | 32GB | 16GB |
| RAM Freq | DDR4 2133 | |
| NIC | Mellanox ConnectX-3 (10Gbps) | |
| OS | CentOS 7.2 | |

**TABLE 1:** Test System Setup

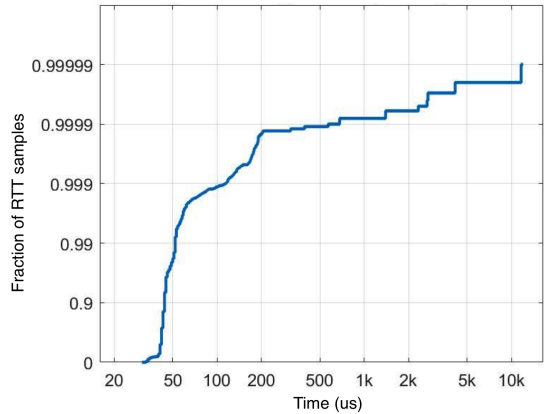

**Fig. 4:** RTT latency for 100K messages of size 1KB for a TCP-IP single-server single-client configuration: Rare tails as high as $11ms$ are observed at the $4^{th}$ and $5^{th}$ 9s

### 5.1 Evaluation Metrics

We evaluate different network protocols based on their $n^{th}$ 9 RTT delays, where n = 3, 4, and 5. n = 3 refers to 99.9% of the total transactions, n = 4 refers to 99.99% of the total transactions, and n = 5 refers to 99.999% of the total transactions. The X-axis for all subsequent graphs represents the RTT latency in microseconds and the Y-axis portrays the fraction of transactions that incur a specific RTT latency.

### 5.2 TCP-IP

We begin our evaluation of the TCP-IP network protocol with the baseline single-server, single-client connection that sends $100,000$ messages of 1KB each. An analysis of the RTT distribution is depicted in Fig. 4. We recognize that 99.9% of messages complete the round trip within $100$ $\mu$s. The more critical and surprising observation are the rare tails as enormous as $11ms$ in the $4^{th}$ and $5^{th}$ 9s for a simple TCP-IP connection. We argue that even the $4^{th}$ and $5^{th}$ 9s can markedly degrade application performance at 100- and 1000-node scale, as a large fraction of such large distributed requests may incur at least one such tail. We note that academic researchers and companies like Amazon and Google have shifted focus from $90^{th}$ and $95^{th}$ percentage tail latencies to $99^{th}$ and $99.9^{th}$ percentage tails in recent studies [4, 16, 7].

Next, we scrutinize the effect of bandwidth utilization on the prevalence and magnitude of the tail. Bandwidth stress tests vary packet sizes from 64B to 512KB, while retaining the single-server, single-client configuration. Trends for the accumulated RTT measurements are illustrated in Fig. 5. As the packet size is increased beyond 1520B, we observe a slight deterioration of the non-tail latency i.e 99.9% of RTT measurements for packet sizes greater than 1520B take longer to complete than 99.9% of RTT measurements for packet sizes lesser than 1520B. This trend is in accordance with the fact that the underlying Ethernet layer splices TCP-IP packets, thereby introducing an additional overhead to median latency. However, the most prominent aspect of our measurement is that significant tail latencies continue to prevail at the $4^{th}$ and $5^{th}$ 9s. Overall, we note that while bigger packet sizes experience a larger median RTT, the payload size has no significant impact on the magnitude or frequency of the tail, indicating that extreme tails are unrelated to payload size or packet splicing.

We then investigate the influence of multiplexing the physical link by initiating multiple connections between endpoints. We vary the number of connections from 1-8, while fixing the packet size at 1KB. We achieve the maximum physical link bandwidth with one TCP-IP connection per thread, hence, we do not study scenarios that multiplex more connections. We again measure $100,000$ RTTs and report the result in Fig. 6. Yet again, we observe that tail latencies as high as $20ms$ continue to dominate at the $4^{th}$ and $5^{th}$ 9s.

Finally, we study the effect of non-blocking TCP-IP system calls on the RTT distribution for a single server-client pair that exchange $100,000$ messages of size 1KB. The RTT distribution is shown in Fig. 7. We observe a median latency smaller than $100$ $\mu$s and the worst-case tail latency is $8ms$. This RTT trend implies that the blocking system calls are not the root cause of extreme TCP-IP latency tails; similar tails occur for both synchronous and asynchronous socket interfaces.

### 5.3 UDP-IP

Analogous to the TCP-IP baseline, we evaluate UDP-IP for a single-server, single-client connection. A breakdown of the packet RTT distribution is shown in

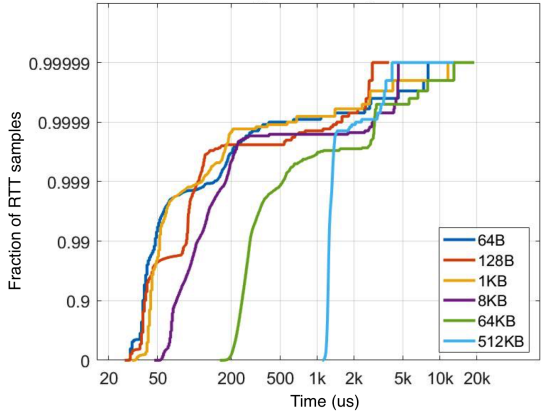

**Fig. 5:** RTT latency of 100K messages of sizes 64B-512KB for a TCP-IP single-server single-client configuration: Rare tails as high as $19ms$ are observed at the $4^{th}$ and $5^{th}$ 9s

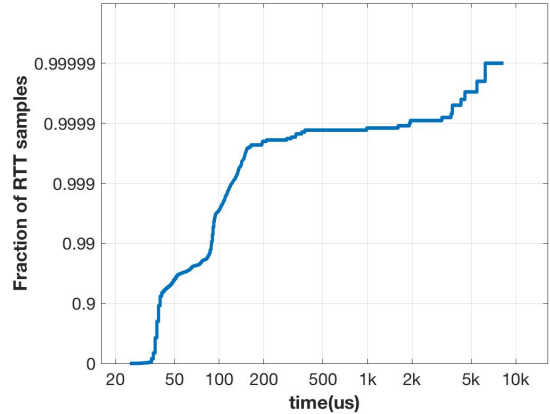

**Fig. 7:** RTT latency for 100K 1KB messages for a single non-blocking TCP-IP server-client connection: Rare tails as high as $8ms$ continue to exist at the $4^{th}$ and $5^{th}$ 9s

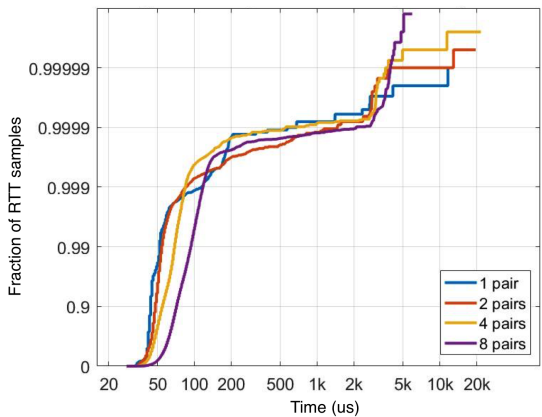

**Fig. 6:** RTT latency for 100K 1KB messages for 1-8 simultaneous TCP-IP server-client connections: Rare tails as high as $20ms$ continue to exist at the $4^{th}$ and $5^{th}$ 9s

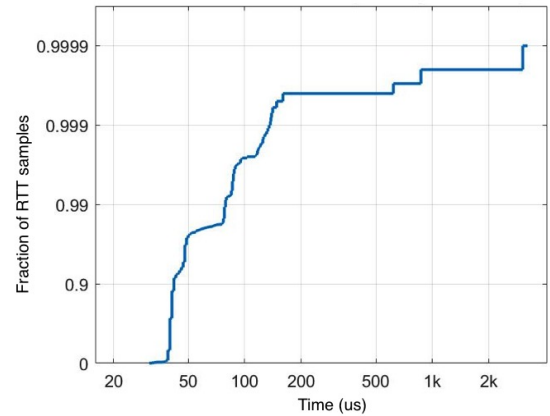

**Fig. 8:** RTT latency for 100K messages of size 1KB for a UDP-IP single-server single-client configuration: Rare tails as long as $2.5ms$ prevail close to the $4^{th}$ 9s

Fig. 8. We notice that, again, 99.9% of messages exhibit an RTT below 100 $\mu s$. More importantly, we see tail latency behavior quite similar to the tails displayed by TCP-IP: tails as prolonged as $2.5ms$ prevail close to the $4^{th}$ 9. The worst-case tail latency of $2.5ms$ is less extreme than the maximum tail latency of $11ms$ observed for our comparable TCP-IP setup. However, this difference is insignificant from an application performance degradation perspective.

We next evaluate the effect of increasing bandwidth utilization on the RTT distribution by increasing packet size and transmitting multiple packets concurrently. We vary packet sizes from 64B-1472B and exchange messages between eight server-client pairs. We do not send packets larger than 1472B as this is the maximum transfer unit size supported by UDP. Our RTT measurements are shown in Fig. 9. We observe tail latencies of at most $11ms$ at the $4^{th}$ 9 and hence conclude that increasing network bandwidth utilization does not exacerbate the frequency of the baseline tails. However, the magnitude of tails (as high as $11ms$) appears to be aggravated relative to the baseline case of $2.5ms$.

### 5.4 RDMA

Finally, to distinguish the impact of software vs. hardware on tails, we perform experiments using RDMA. Payloads varying from 64B-8KB are read/written to/from the remote memory, in a single-server, single-client configuration. The experiment measures $100,000$ data ping-pongs for every packet size.

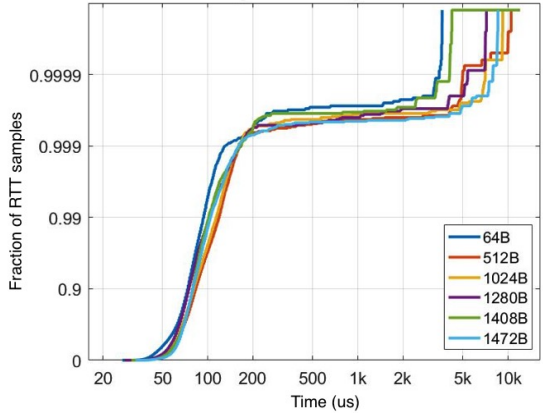

**Fig. 9:** RTT latency for 100K messages of sizes 64B-1472B for 8 UDP-IP server-client pairs: Rare tails as high as $11ms$ observed at the $4^{th}$ 9

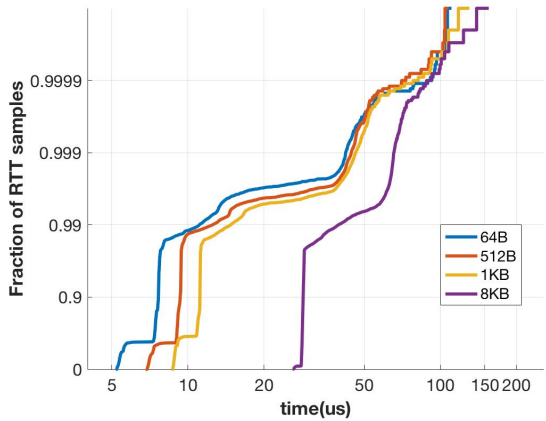

**Fig. 10:** RDMA round trip read latency for 100K messages of sizes 64B-8KB: No extreme tail latencies observed

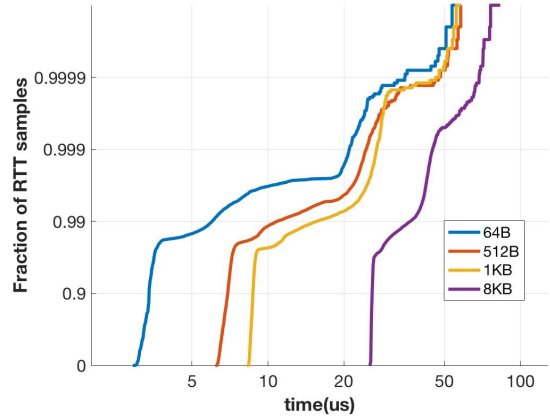

**Fig. 11:** RDMA round trip write latency for 100K messages of sizes 64B-8KB: No extreme tail latencies observed

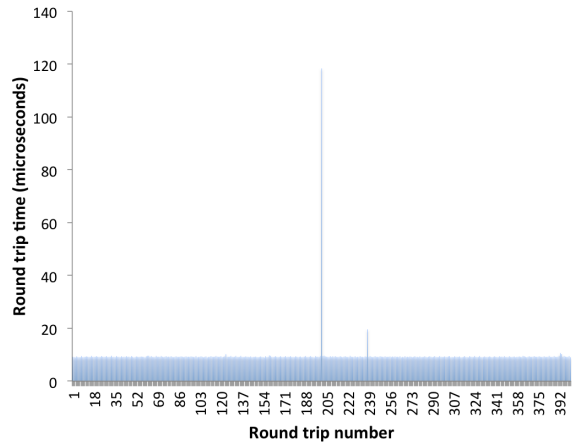

**Fig. 12:** RTT for some RDMA read accesses: High latency samples are rare and distinct

modest and provide strong evidence that the kernel networking stack is the root cause of extreme tails.

RTT latencies observed for RDMA reads from the remote memory are illustrated in Fig. 10. Fig. 11 indicates the RTT distribution for RDMA writes to the remote memory. A majority of RDMA read/write accesses experience an RTT latency around 10 $\mu$s. More importantly, the absence of extreme tail latencies in RDMA read/write RTTs is clear in these graphs. The worst-case RTT (for RDMA read) of 150 $\mu$s is 15 times the median, an order of magnitude smaller than the ratio of worst-case tail to median latency for TCP-IP and UDP-IP round-trips. Moreover, the absolute value of the worst-case RTT for TCP-IP/UDP-IP ($11ms/2.5ms$) is greater than the RDMA worst-case RTT by more than one order of magnitude. The small ratio of the worst-case tail to median latency and the small absolute value of the worst-case tail latency exhibited by RDMA, implies that RDMA tails are modest and provide strong evidence that the kernel networking stack is the root cause of extreme tails.

Although its tails are less extreme, 0.1% of RTTs are still as high as 150 $\mu$s for RDMA accesses. On interesting question is whether high-latency RTTs occur independently (suggesting a root cause that affects only a single packet, interrupt delivery, task switch, etc.) or in bursts (suggesting a sustained disruption, such as CPU transitioning to a lower-performance power mode). Fig. 12 represents a snapshot of the $100,000$ RTT samples collected for the RDMA read operation. A high RTT of 120 $\mu$s observed only for the $196^{th}$ round-trip, while neighboring RTTs incur median latencies. Hence, we conclude that RDMA latency tails are primarily due independent events affecting an individual packet.

## 6 DISCUSSION

By the process of elimination, we arrive at our key conclusion that latency bottlenecks in the commonly used network protocol such as TCP-IP are due to the influence of the software protocol stack in the operating system kernel. We consider and reject several alternative hypotheses:

**Eliminating extrinsic network parameters**
Our experiment design eliminates the effect of less innate network aspects like network switch congestion, packet scheduling, and load balancing on performance variations. Configuring an isolated network by simply connecting two machines with a physical link precludes these extrinsic sources of tail latency; any tail latencies observed in our experiments must be due to behavior in the end systems, their network interfaces, or due to packet corruption on the wire (which we deem unlikely).

**Eliminating TCP-IP complexities.**
TCP-IP protocol complexities like windowing, packet ordering, and the three-way handshake operations may potentially deteriorate application performance by introducing long tail latencies. Our UDP-IP experiments confirm that the median RTT of $100$ $\mu s$ is equivalent to that of TCP-IP. Moreover, tail latencies as high as $2.5ms$ exhibited by UDP-IP, are comparable to the TCP-IP tail manifestations ( $11ms$). The slight difference in the magnitude of rare tails seen by these protocols is insignificant from a performance degradation perspective. Hence, for all practical purposes, extreme tail behavior under both TCP-IP and UDP-IP is the same. As the UDP-IP protocol is devoid of protocol complexities present in TCP-IP and UDP-IP still exhibits a significant tail, we conclude that TCP's additional features are not the root cause of the extreme latency tails we observe.

**Eliminating non-OS bottlenecks.**
When extrinsic network parameters and TCP-IP protocol intricacies are ruled out as probable miscreants that degrade application performance, we are left with two other potential tail-inducing candidates: (1) non-OS protocol components and (2) the design of the software-based operating system stack invoked by the TCP-IP and UDP-IP protocols.

RDMA performs read/write actions directly on the remote memory, thereby reducing data transfer latency of $100$ $\mu s$ (as is the case with TCP-IP/UDP-IP) to $10$ $\mu s$. The RDMA protocol exchanges memory region keys directly between the server and client via the physical link to send/receive data. Therefore, the RDMA protocol design does not involve any operating system actions during a data transfer; latency overhead is incurred solely by the RDMA network adapter hardware performing send/receive operations on remote memory locations or the user-mode software that initiates the transfer. As RDMA RTTs are at most $150$ $\mu s$, more than an order of magnitude smaller than the TCP-IP and UDP-IP tails, it is justifiable to rule out the user-mode client software as the root cause of extreme tail latency.

Therefore, by the process of elimination, we deduce that throughput bottlenecks in the TCP-IP/UDP-IP protocol stack within the operating system kernel are primarily responsible for extreme tail latencies. Apart from our experiments, this conclusion is supported by the fact that TCP-IP/UDP-IP protocols were originally designed for a generation of computation where processing times were orders of magnitude greater than today's sub-millisecond delays. Therefore, the millisecond tails generated by these protocols were inconsequential in the past, as they were hidden by computation latencies. However, as contemporary computation operates at GHz frequencies, the legacy TCP-IP/UDP-IP protocol latencies can no longer be hidden by compute time and therefore manifest as glaring millisecond tails.

Furthermore, it is imperative to investigate the bottlenecks imposed by individual components of the TCP-IP/UDP-IP software stack. The TCP-IP/UDP-IP software stack is designed in a manner that requires the data and commands to be read from the network adapter into system memory before they can be processed. Therefore, three memory copies are necessary for each TCP-IP/UDP-IP packet: (1) to read the data and command from the NIC to the IP streaming area (2) to read and process the command; and (3) to write and move the data to the user. Reading/writing to/from these three memory copies creates a significant overhead as it invokes up to three operating system interrupts along with data movement [19]. We suspect that the dominant tail-inducing bottleneck in these protocols is the need for frequent operating system interrupts. Redesigning the network protocol components such that incessant system calls and interrupts are eradicated could potentially curtail long tails. We propose to identify the contribution of each element in the TCP-IP/UDP-IP software handler to the overall latency distribution, as part of future work.

## 7 CONCLUSION

Tail latency is a significant source of performance degradation in warehouse-scale computing. The network layer is a key contributor to large tail latencies in OLDI applications like web search and online retailing. Network protocol design is a foundational aspect of the network layer and therefore we deconstructed the tail-at-scale effect across the most commonly used network protocols: TCP and UDP over IP.

We identified surprising rare tails of significant magnitude in TCP-IP message exchanges. We performed an analysis of various factors that could inhibit application performance and identified the true culprit, the OS protocol stack, by the process of elimination. Increasing bandwidth utilization did not worsen the original tail latencies. The UDP-IP protocol did not mitigate the tail effect and therefore additional TCP-IP protocol complexities were ruled out as probable tail-generating candidates. Finally, the fact that the RDMA protocol did not exhibit a similar magnitude of tail latencies led us to conclude that operating system networking protocol stack is the primary source of tail latencies in the TCP/UDP over IP protocols.

## 8 ACKNOWLEDGEMENT

This work was supported in part by NSF Grant IIS-1539011 and gifts from Intel.

We also thank Aasheesh Kolli, Neha Agarwal, Vaibhav Gogte and the anonymous referees for their insightful feedback on drafts of this work.

## REFERENCES

[1] Chang-Hong Hsu et al. "Adrenaline: Pinpointing and Reining in Tail Queries with Quick Voltage Boosting". In: *Proc. of the Symp. on High Performance Computer Architecture*. 2015.

[2] Jialin Li et al. "Tales of the Tail: Hardware, OS, and Application-level Sources of Tail Latency". In: *Proc. of the ACM Symposium on Cloud Computing*. 2014.

[3] Timothy Zhu et al. "PriorityMeister: Tail Latency QoS for Shared Networked Storage". In: *Proc. of the ACM Symposium on Cloud Computing*. 2014.

[4] Jeffrey Dean and Luiz André Barroso. "The Tail at Scale". In: *Communications of the ACM* 56 (2013), pp. 74–80.

[5] Virajith Jalaparti et al. "Speeding Up Distributed Request-response Workflows". In: *Proc. of ACM SIGCOMM*. 2013.

[6] Rajesh Nishtala et al. "Scaling Memcache at Facebook". In: *10th USENIX Symposium on Networked Systems Design and Implementation*. 2013.

[7] Yunjing Xu et al. "Bobtail: Avoiding Long Tails in the Cloud". In: *10th USENIX Symposium on Networked Systems Design and Implementation*. 2013.

[8] Mohammad Alizadeh et al. "Less Is More: Trading a Little Bandwidth for Ultra-Low Latency in the Data Center". In: *9th USENIX Symposium on Networked Systems Design and Implementation*. 2012.

[9] Chi-Yao Hong, Matthew Caesar, and P. Brighten Godfrey. "Finishing Flows Quickly with Preemptive Scheduling". In: *Proc. of the ACM SIGCOMM 2012 Conference on Applications, Technologies, Architectures, and Protocols for Computer Communication*. 2012.

[10] Balajee Vamanan, Jahangir Hasan, and T. N. Vijaykumar. "Deadline-Aware Datacenter TCP (D2TCP)". In: *Proc. of ACM SIGCOMM*. 2012.

[11] David Zats et al. "DeTail: Reducing the Flow Completion Time Tail in Datacenter Networks". In: *Proc. of ACM SIGCOMM*. 2012.

[12] David Meisner et al. "Power Management of Online Data-intensive Services". In: *Proc. of the International Symposium on Computer Architecture*. 2011.

[13] Christo Wilson et al. "Better Never Than Late: Meeting Deadlines in Datacenter Networks". In: *Proc. of ACM SIGCOMM*. 2011.

[14] Mohammad Alizadeh et al. "Data Center TCP (DCTCP)". In: *Proc. of ACM SIGCOMM*. 2010.

[15] T. Hoff. *Latency is Everywhere and it Costs You Sales - How to Crush it*. 2009. URL: http://highscalability.com/blog/2009/7/25/latency-iseverywhere-and-it-costs-you-sales-how-to-crush-it.html.

[16] Giuseppe Decandia et al. "Dynamo: amazon's highly available key-value store". In: *ACM Symposium on Operating Systems Principles*. 2007.

[17] Ron Kohavi, Randal M. Henne, and Dan Sommerfield. "Practical Guide to Controlled Experiments on the Web: Listen to Your Customers Not to the Hippo". In: *Proc. of the International Conference on Knowledge Discovery and Data Mining*. 2007.

[18] Erich M. Nahum, John Tracey, and Charles P. Wright. "Evaluating SIP Server Performance". In: *Proc. of the SIGMETRICS*. 2007.

[19] Pavan Balaji, Hemal Shah, and D.K. Panda. "Sockets vs RDMA Interface over 10-Gigabit Networks: An In-depth analysis of the Memory Traffic Bottleneck". In: *RAIT workshop '04*. 2004.

[20] Luiz André Barroso, Jeffrey Dean, and Urs Hölzle. "Web Search for a Planet: The Google Cluster Architecture". In: *IEEE Micro*. Vol. 23. 2. 2003, pp. 22–28.

[21] Anna Bouch, Nina Bhatti, and Allan Kuchinsky. "Quality is in the Eye of the Beholder: Meeting Users' Requirements for Internet Quality of Service". In: *ACM Conference on Human Factors and Computing Systems*.

[22] J. Rothschild. *High performance at massive scale: Lessons learned at facebook*. URL: mms://video-jsoe.ucsd.edu/calit2/JeffRothschildFacebook.wmv.